\documentclass[twocolumn,showpacs,preprintnumbers,amsmath,amssymb]{revtex4}
\usepackage{amsmath,amssymb,graphics,epsfig,subfigure}
\usepackage{color}

\usepackage{color}

\begin{document}
\renewcommand{\baselinestretch}{1.3}

\title{Static spheres around spherically symmetric black hole spacetime}

\author{Shao-Wen Wei$^{1,2,3}$ \footnote{weishw@lzu.edu.cn},
Yu-Peng Zhang$^{1,2,3}$ \footnote{zyp@lzu.edu.cn},
Yu-Xiao Liu$^{1,2,3}$ \footnote{liuyx@lzu.edu.cn},
Robert B. Mann$^{4}$ \footnote{rbmann@uwaterloo.ca}}

\affiliation{$^{1}$ Key Laboratory of Quantum Theory and Applications of MoE, Lanzhou University, Lanzhou 730000, China\\
$^{2}$Lanzhou Center for Theoretical Physics, Key Laboratory of Theoretical Physics of Gansu Province, School of Physical Science and Technology, Lanzhou University, Lanzhou 730000, People's Republic of China,\\
$^{3}$Institute of Theoretical Physics $\&$ Research Center of Gravitation,
Lanzhou University, Lanzhou 730000, People's Republic of China,\\
$^{4}$ Department of Physics \& Astronomy, University of Waterloo, Waterloo, Ont. Canada N2L 3G1}

\begin{abstract}
Unique features of particle orbits produce novel signatures of gravitational observable phenomena, and are quite useful in testing compact astrophysical objects in general relativity or modified theories of gravity. Here we observe a representative example that a static, spherically symmetric black hole solution with nonlinear electrodynamics admits static points at finite radial distance. Each static point thus produces a static sphere, on which a massive test particle can remain at rest at arbitrary latitudes with respect to an asymptotic static observer. As a result, the well-known static Dyson spheres can be implemented by such orbits. More interestingly, employing a topological argument, we disclose that stable and unstable static spheres (if they exist) always come in pairs in an asymptotically flat spacetime. In contrast to this, the counterpart naked singularity has one more stable static sphere than the unstable one. Our results have potential applications in testing black holes in standard Maxwell and nonlinear electrodynamics, as well as in uncovering the underlying astronomical observation effects in other gravitational theories beyond general relativity.
\end{abstract}



\maketitle

\section{Introduction}

Exploring the nature of a black hole or spacetime in both weak and strong gravity regimes largely relies on the characteristic geodesic motions of test particles. Some characteristic phenomena, such as the ringdown of black hole binaries \cite{Abbott} and shadows \cite{Akiyama1,Akiyama2}, can also be understood by the circular orbits of photons or massive particles. Further study also shows that semi orbits, pointy petal orbits, and static light points appear in rotating boson stars, hairy black holes, wormholes \cite{Grandclement,Grould,Teodoro,Yazadjiev}, and higher-dimensional rotating black holes \cite{Gibbons,Herdeiro,Diemer}, expecting to have potential signatures on observable effects. Recent developments can also be found in Refs. \cite{Delgado,Lehebel}.

It is generally believed that the test massive particles with vanishing angular momentum cannot remain at rest at finite radial distance around a black hole in general relativity. In particular, due to frame dragging, they will orbit a rotating black hole. However by tuning the angular momentum of the counter-rotating particles a static point in the equatorial plane can be obtained \cite{Collodel}. Due to the symmetry of the spacetime, there will be a special orbit, the ring of static points, on which particles initially at rest remain at rest with respect to an asymptotic static observer. Such particular orbits are found in rotating non-Kerr black holes. Therefore observing static, or quasi-static, phenomena could offer support for the presence of compact objects that differ from Kerr black holes.

It is natural to ask whether a similar static sphere exists for a non-rotating black hole. If so, any massive particle will remain rest at arbitrary latitudes. Thus a thin static shell can be formed only under the gravitation interaction. Consequently this mimics an actual and rigid Dyson sphere. Proposed by Freeman Dyson \cite{Dyson}, this object was a shell initially constructed around a star and designed to exploit all the infrared radiation energy of stellar sources. Its greatest advantage is that it eliminates the inner stress density of the shell or the required extremely high elastic modulus of the shell material \cite{Papagiannis,Wright}.

A characteristic feature of a static point is that the angular velocity vanishes at radius $r_{\text{sp}}$
\begin{eqnarray}
 \Omega(r_{\text{sp}})=0,\label{ome}
\end{eqnarray}
which we required is independent of the polar and azimuth angles in a non-rotating black hole background. In general relativity, minimally coupled to Standard Model matter, such a static sphere is absent, even by decreasing the black hole spin accompanied by a ring of static point.

Here we demonstrate that a simple generalization of electromagnetism, referred to as quasi-topological electromagnetism \cite{Weiladf}, has a dyonic black hole solution that yields the necessary conditions for a Dyson sphere without the necessity of extremality. This theory is a reasonable physical competitor to standard electromagnetism, since its basic effects are not manifest in Earth-based lab experiments \cite{Weiladf}. Via a topological argument, we also show, for an asymptotically flat spacetime, that
 stable and unstable static spheres (if they exist) always come in pairs.

The present work is organized as follows. In Sec. \ref{ssaso}, we start with a static and spherically symmetric black hole. The condition for the static spheres are given. Then taking the quasi-topological electromagnetic black hole as an example, we clearly exhibit the static spheres and straight orbits. Of particular interest, in Sec. \ref{mpss}, the number of the static spheres are studied by using the topological approach. Finally, we summarize and discuss our results in Sec. \ref{summary}.

\section{Static spheres and straight orbits}
\label{ssaso}

Let us briefly review the geodesic equations and analyze the necessary conditions for the existence of the static spheres. The line element of a static, spherically symmetric black hole is
\begin{eqnarray}
 ds^2=-f(r)dt^2+\frac{1}{g(r)}dr^2+r^2(d\theta^2+\sin^2\theta d\phi^2),\label{ele}
\end{eqnarray}
where the metric functions $f(r)$ and $g(r)$ dependent only on the radial coordinate $r$. Without loss of generality, we focus on equatorial geodesics with $\theta=\pi/2$. The metric (\ref{ele}) has two Killing vectors $\xi^{\mu}=(\partial_{t})^{\mu}$ and $\psi^{\mu}=(\partial_{\phi})^{\mu}$, from which follow two conserved quantities
\begin{eqnarray}
 -E=g_{tt}\dot{t}, \quad l=g_{\phi\phi}\dot{\phi}, \label{conservedL}
\end{eqnarray}
which are respectively the energy and orbital angular momentum per unit mass of a massive test particle; the dots denote the derivative with respect to an affine parameter. For a massive test particle, we have $g_{\mu\nu}\dot{x}^{\mu}\dot{x}^{\nu}=-1$. Solving the equation of geodesics, one easily obtains the radial motion of a massive particle
\begin{eqnarray}
 \dot{r}^2+V_{\text{eff}}=0,
\end{eqnarray}
where the effective potential is given by $V_{\text{eff}} =g(r) \left(1+l^2/r^2\right)  - g(r)E^2/f(r)$. For a distant static observer at rest, the angular velocity of a massive test particle circling the black hole can be solved as
\begin{eqnarray}
 \Omega=\frac{f(r)l}{r^2E}.
\end{eqnarray}
Note that outside the black hole horizon, $f(r)>0$ and $g(r)>0$. So vanishing $\Omega$ requires $ l=0$ as expected. Then the reduced effective potential reads
\begin{eqnarray}
 \mathcal{V}_{\text{eff}}=g(r)\left(1-\frac{E^2}{f(r)}\right).\label{adad}
\end{eqnarray}
The static sphere is also a specific circular orbit, which satisfies $\mathcal{V}_{\text{eff}}=\mathcal{V}'_{\text{eff}}=0$, where the prime denotes the derivative with respect to $r$. As a result, the following conditions should be satisfied:
\begin{eqnarray}
 E=\sqrt{f(r_{\text{sp}})},\quad f'(r_{\text{sp}})=0,\label{ccc}
\end{eqnarray}
where $r_{\text{sp}}$ is the radius of the static sphere. Note that $f''(r_{\text{sp}})>0$ and $<0$ correspond to stable and unstable static spheres, respectively, and so the former is of particular interest. It is also worth to point out that the above analysis on the static sphere is general and can be applied to any static, spherically symmetric black hole. After a simple algebraic calculation, one easily finds that there is no static sphere outside the horizon of a Reissner-Nordstr\"{o}m black hole in standard Maxwell electrodynamics. However, when the nonlinear electrodynamic terms are included in, the static sphere will be present.

To illustrate such result, let us consider the dyonic black hole given in Ref. \cite{Liu2020}. A quasi-topological electromagnetic action term is included in
\begin{eqnarray}
 S=\frac{1}{16\pi}\int\sqrt{-g}d^4x\left(R-\alpha_1F^2-\alpha_2
   \left((F^2)^2-2F^{(4)}\right)\right),\nonumber\\
 \label{action}
\end{eqnarray}
where the field strength is $F^2=F^{\mu\nu}F_{\mu\nu}$ and $F^{(4)}=F^{\mu}_{\;\;\nu}F^{\nu}_{\;\;\rho}F^{\rho}_{\;\;\sigma}F^{\sigma}_{\;\;\mu}$. The coupling parameters $\alpha_1$ and $\alpha_2$ are for the standard Maxwell and quasi-topological electromagnetic actions, respectively, with $\alpha_1=1$ and $\alpha_2=0$, yielding standard Maxwell theory. Under the ansatz of global polarization, this quasi-topological term has no influence on the Maxwell equation and the energy-momentum tensor. Note that the $\alpha_2$ term contributes to photon-photon scattering; from accelerator experiments, $\alpha_2 \lesssim \frac{\hbar e^4}{360 \pi^2m_{e}^4c^7}$. For a black hole of mass $M$, $\alpha_2/M^2$ is dimensionless in geometric units $c=\hbar=k_B=G=1$; to obtain $\alpha_2/M^2<1$, $M > 10^4M_{\odot}$. Hence this solution can be applied to supermassive black holes located in the centers of galaxies. This bound could be further reduced by the cancellation of the two $\alpha_2$-dependent terms in \eqref{action}.

The static spherically symmetric black hole solution to the field equations following from \eqref{action} is \cite{Liu2020}
\begin{eqnarray}
 f(r)=g(r)&=&1-\frac{2M}{r}+\frac{\alpha_1 p^2}{r^2}\nonumber\\
 &+&\frac{q^2}{\alpha_1r^2}~_{2}F_1
    \left[\frac{1}{4},1; \frac{5}{4}; -\frac{4p^2\alpha_2}{r^4\alpha_1}
    \right],\label{fr}
\end{eqnarray}
with $~_{2}F_1$ the hypergeometric function. The electric and magnetic charges are $q$ and $p/\alpha_1$, respectively. For this matter sector, the null, weak, and dominant energy condition are held when both $\alpha_1$ and $\alpha_2$ are positive, while the strong energy condition is violated. We note that other forms of matter, such as a scalar field with a positive potential, along with most inflationary models, also violate this condition. In this sense the quasi-topological term behaves like dark energy.

Most intriguingly, we have a class of black hole solutions respecting standard energy conditions that exhibits interesting results. The chaos bound does not seem to be universally satisfied \cite{Lei}, and echoes can also naturally emerge without placing a hard wall near the horizon \cite{Huang}. From the viewpoint of black hole thermodynamics, a triple point phase structure is present, indicating a rich underlying microstructure \cite{Li}. In particular, in certain parameter regions, for example, $\alpha_1=1,\;\alpha_2/M^2=2.76,\;p/M=0.14,\; q/M=1.02$, there are four black hole horizons and three photon spheres, indicating an interesting spacetime structure. Such intriguing properties will cast potential imprints on black hole image and gravitational waves.

\begin{figure}
\center{
\includegraphics[width=6cm]{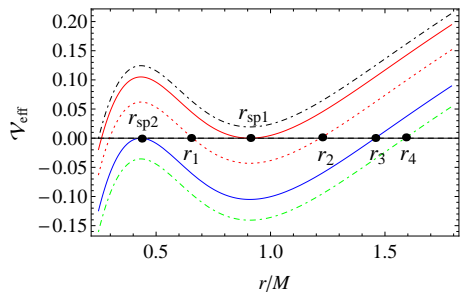}}
\caption{The reduced effective potential (\ref{adad}) for the dyonic black hole. The energy of the massive test particle is set as $E$=0.01, 0.1393 ($E_{\text{sp1}}$), 0.25, 0.3528 ($E_{\text{sp2}}$), and 0.4 from top to bottom. The stable and unstable static spheres are located at $r_{\text{sp1}}/M=0.91$ and $r_{\text{sp2}}/M=0.43$, respectively.}\label{ppVeff}
\end{figure}

Plotting the reduced effective potential (\ref{adad}) in Fig. \ref{ppVeff}, we observe one stable and one unstable static sphere at $r_{\text{sp1}}$ and $r_{\text{sp2}}$ for fixed energy $E_{\text{sp1}}$ and $E_{\text{sp2}}$, respectively. For each energy $E\in$ ($E_{\text{sp1}}$, $E_{\text{sp2}}$), there is a bound orbit. A neutral massive test particle will undergo straight back-and-forth motion between two turning points, for example $r_1$ and $r_2$ for $E$=0.25. Note that there is no angular motion due to the vanishing angular velocity.

\begin{figure}
\center{\subfigure[]{\label{Veffa}
\includegraphics[width=3.5cm]{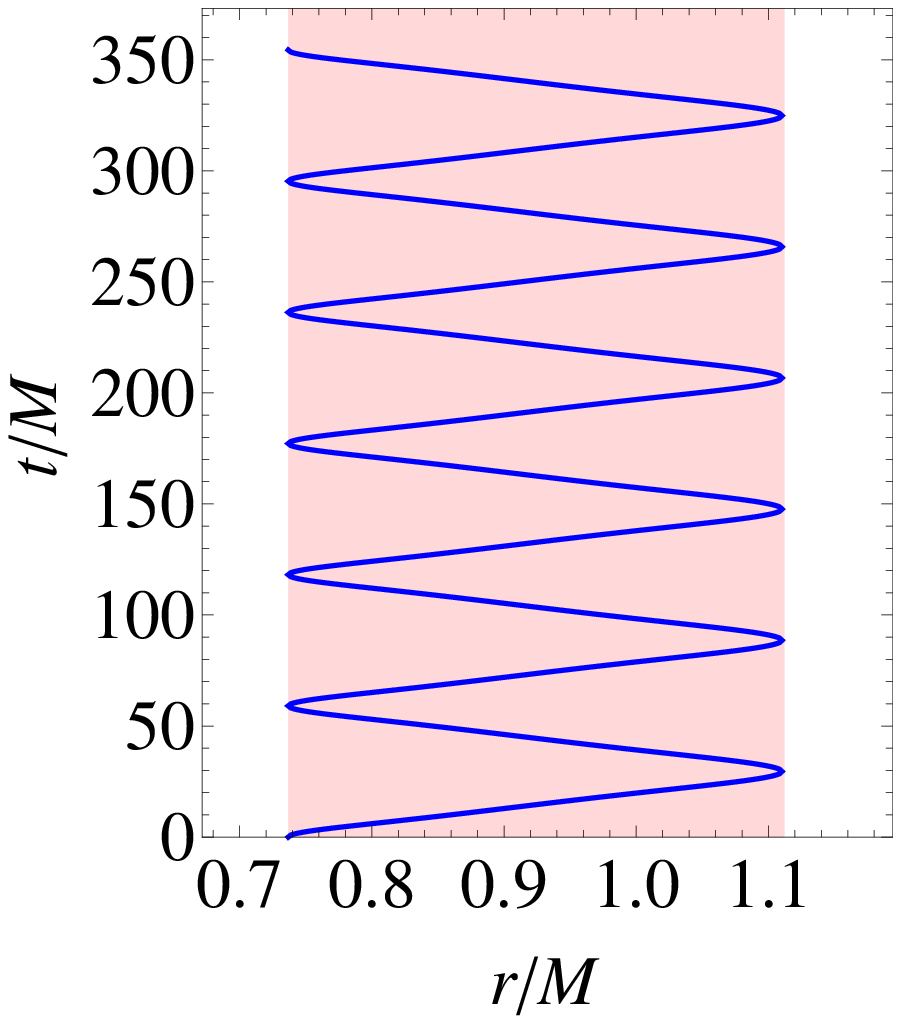}}
\subfigure[]{\label{Veffb}
\includegraphics[width=3.5cm]{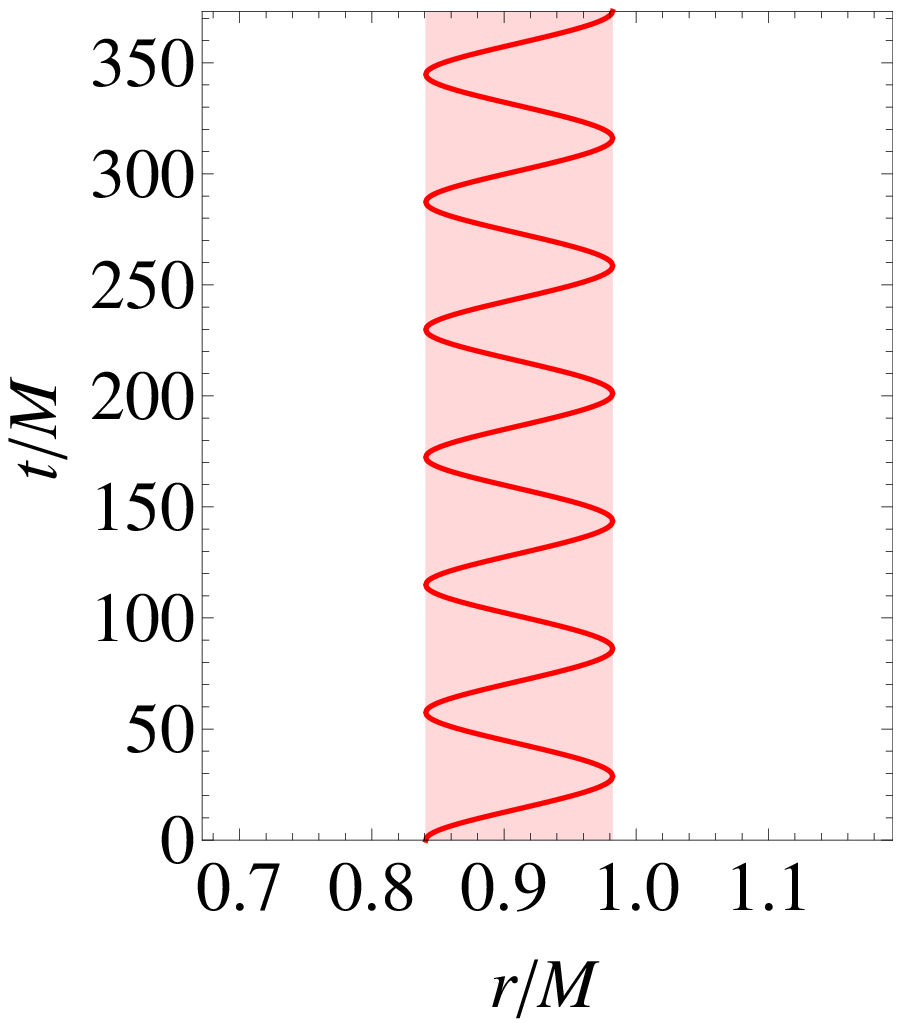}}\\
\subfigure[]{\label{Veffc}
\includegraphics[width=3.5cm]{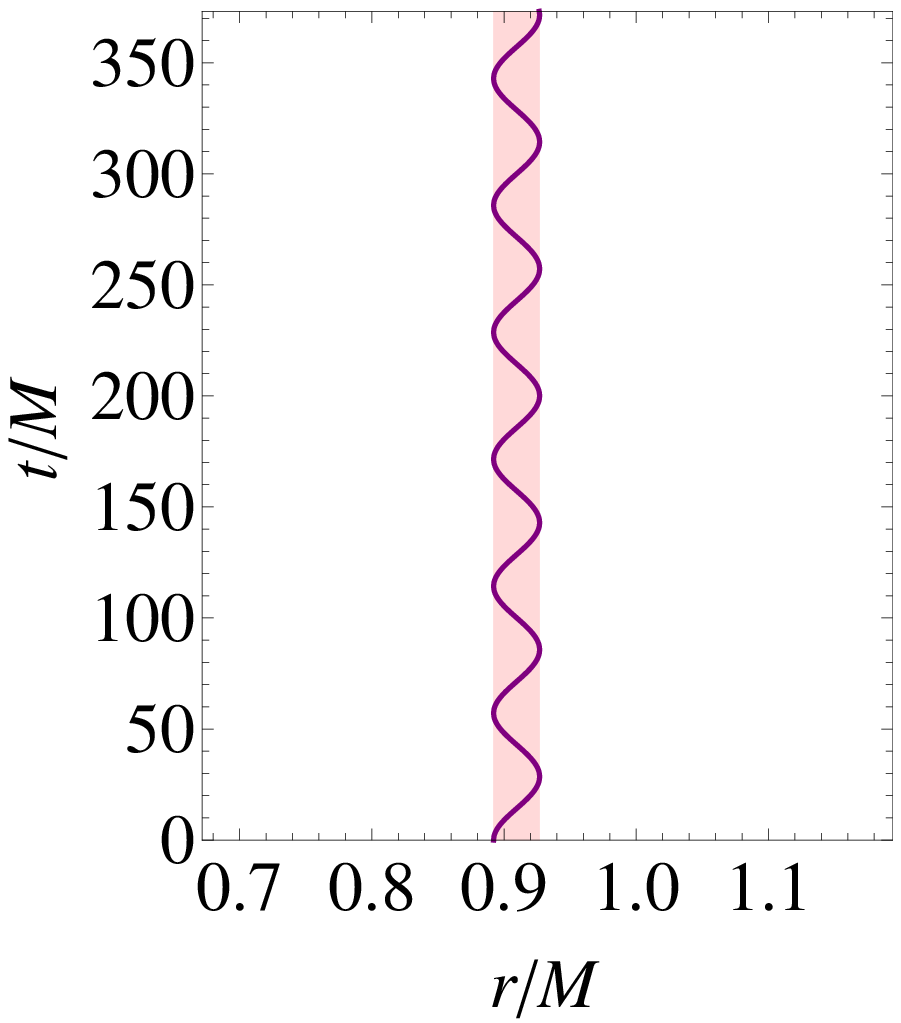}}
\subfigure[]{\label{Veffd}
\includegraphics[width=3.5cm]{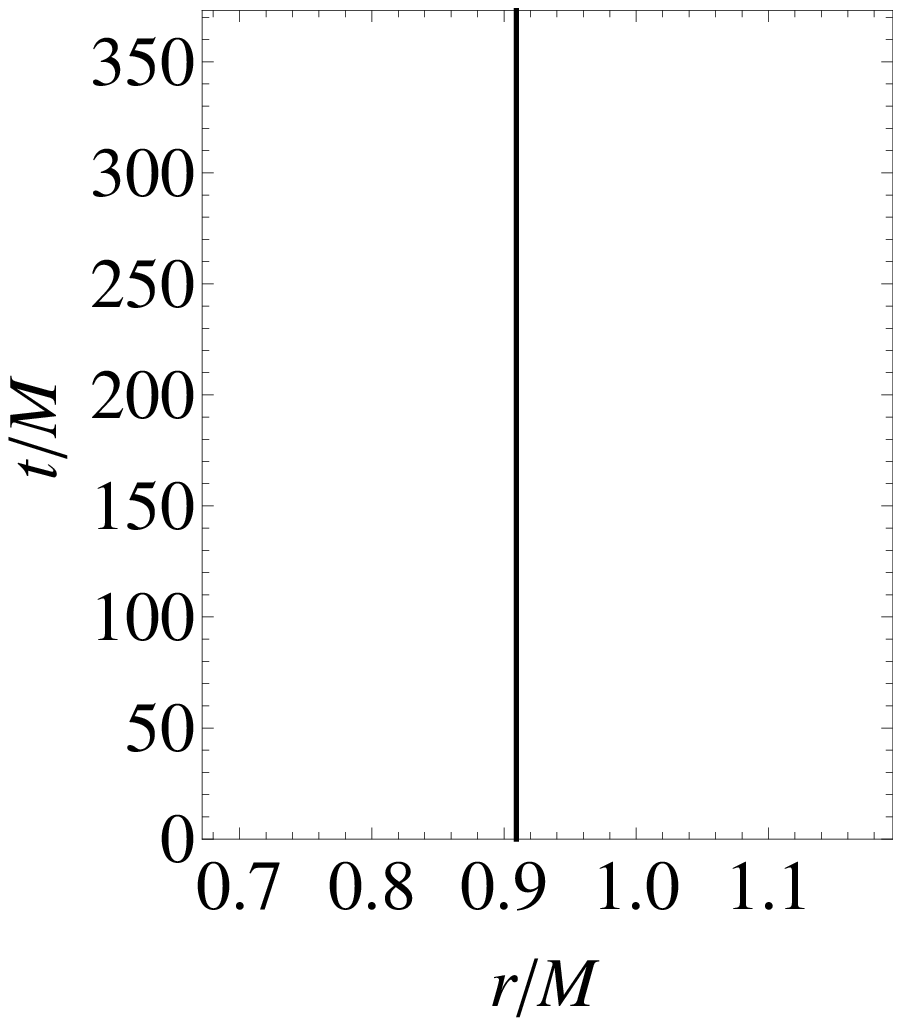}}}
\caption{Radial motion of the massive test particle. (a) $E$=0.2. (b) $E$=0.15. (c) $E$=0.14. (d) $E$=$E_{\text{sp1}}$. When $E$=$E_{\text{sp1}}$, the radial motion is just a horizontal line denoting a static orbit with vanishing angular momentum.}\label{pprt}
\end{figure}

Taking energies to be $E$=0.2, 0.15, 0.14, and $E_{\text{sp1}}=0.1393$, we exhibit the radial motion of the particle starting at the small turning points in the $r-t$ plane in Fig. \ref{pprt}. Obviously, when the energy decreases and tends to $E_{\text{so1}}$, the radial region of the motion narrows. In particular, when $E=E_{\text{sp1}}$ shown in Fig. \ref{Veffd}, the radial distance remains unchanged with coordinate time, indicating there is a static sphere at $r=r_{\text{sp1}}$.

In summary, we observe two static spheres -- one stable and one unstable -- for the static, spherically symmetric dyonic black hole. Thus a static Dyson-like sphere can be appropriately constructed. Actually, the existence of the static spheres is due to the effective repulsion produced by the nonlinear electrodynamics behaved like dark energy, which is quite different from the case for the rotating black holes balanced by the nonvanished angular momentum \cite{Collodel}.

\section{Number of static spheres}
\label{mpss}

As shown above, there is one stable and one unstable static spheres. We now determine the number of static spheres for a general static, spherically symmetric black hole \eqref{fr}, providing us with universal information not requiring refer to a specific spacetime. This approach was used to show that light rings are an intrinsic structure of spacetime, independent of the photon \cite{Cunhaa,Cunhab} (see also Refs. \cite{Hod2018plb,Hod2018epjc,Weisw,Guo,Ghosh,Junior,Sanchis,Hod2023prd}), with at least one ring being radially unstable. This is quite different from timelike circular orbits, which closely depend on the energy and angular momentum of the test particle. Quite remarkably, a well-behaved topology characterizing equatorial timelike circular orbits can be constructed \cite{Weila}. In what follows, we mainly focus on the topological properties for a static sphere without orbital angular momentum.

In order to satisfy the conditions for a static sphere given in (\ref{ccc}), we construct the vector $\phi$=($\phi^r$, $\phi^{\theta}$) with
\begin{eqnarray}
 \phi^r=\frac{\partial f(r)}{\partial r},\quad
 \phi^{\theta}=-\frac{\cos\theta}{\sin^2\theta}.\label{vect}
\end{eqnarray}
Note that $\phi^{\theta}$ is an auxiliary term given in Ref. \cite{Weiladf}, which allows us to explicitly show the direction of the vector in a two dimensional plane. The auxiliary term can also be selected as some other smooth and continuous function of $r$ and $\theta$ without introducing extra zero point of $\phi$. Obviously a static sphere is exactly located at a zero point of $\phi$. For any given static sphere we can calculate its winding number, where a positive (negative) value signifies a stable (unstable) static sphere. However, here we are concerned with the sum of the winding numbers, namely the total topological number $W$ of the static spheres corresponding to the following topological current
\cite{Duan}
\begin{eqnarray}
 j^{\mu}=\frac{1}{2\pi}\epsilon^{\mu\nu\rho}\epsilon_{ab}\partial_{\nu}n^{a}\partial_{\rho}n^{b},
 \quad \mu,\nu,\rho=0,1,2,
\end{eqnarray}
where $\partial_{\nu}=\frac{\partial}{\partial x^{\nu}}$ and $x^{\nu}=(\tau,~r,~\theta)$ and the unit vector is defined as $n^a=\frac{\phi^a}{||\phi||}$ ($a=1, 2$). The parameter $\tau$ is a time control parameter. It is not hard to check that this current is conserved: $\partial_{\mu}j^{\mu}=0$.

\begin{figure}
\center{
\includegraphics[width=6cm]{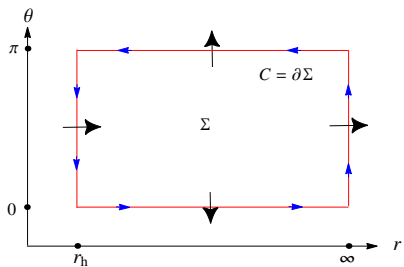}}
\caption{Representation of the vector direction along the boundary. The black arrows represents the direction of the vector, and the closed rectangular loop denotes the complete boundary of the parameter space.}\label{ppWinding}
\end{figure}

To determine $W$, we need to examine the behavior of the vector $\phi$ at the boundary of the $r-\theta$ plane. A simple calculation shows that at $\theta$=0 and $\pi$, the direction of $\phi$ is outwards. On the other hand, near the horizon $r=r_{\text{h}}$, $f(r_{\text{h}})=0$ and $f(r>r_{\text{h}})>0$; thus $\phi^{r}$ is positive, which indicates the direction of $\phi$ is rightward in the plane, ignoring the specific values of $\phi^{\theta}$. At large $r$, for an asymptotically flat black hole, it is easy to obtain $f'(r\rightarrow\infty)>0$. This suggests that the direction of $\phi$ is also towards the right. For clarity, we sketch the direction of the vector (\ref{vect}) in Fig. \ref{ppWinding}. The closed rectangular loop denotes the complete boundary of the parameter space. The black arrows represents the direction of the vector. It is clear that that the direction of the vector changes at different segments of the boundary. Going counterclockwise along the loop once, the direction of the vector does not make one completely loop. Thus, we easily obtain the total topological number
\begin{eqnarray}
 W=\sum_i w_i=0,
\end{eqnarray}
where $w_i$ denotes the winding number of the $i$-th zero point of $\phi$. Since our result is universal, for an arbitrary asymptotically flat, static, spherically symmetric black holes, it strongly suggests that if static spheres exist, they always come in pairs. It is easy to find that the radial stable and unstable static spheres has winding number $w$=1 and -1. So if one is radial stable, another must be unstable. We shall see that this result changes for naked singularities.

\begin{figure}
\center{
\includegraphics[width=6cm]{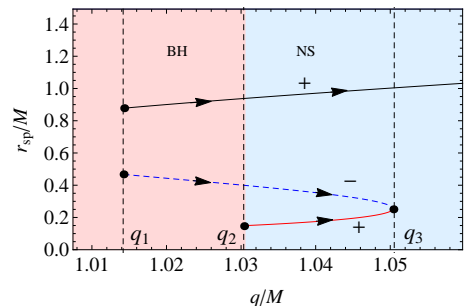}}
\caption{The radius of the static sphere as a function of the time control parameter $q$. ``BH" and ``NS" are for the black hole and naked singularity, respectively. The signs ``+" and ``-" denote the positive and negative winding numbers, respectively. Three characteristic charges are localized at $q_1/M$=1.0143, $q_2/M$=1.0305, and $q_3/M$=1.0504. The black arrows indicate the increase of the time control parameter.}\label{pprqso}
\end{figure}

As shown in Ref. \cite{Weisw}, the black hole charge can be treated as a time control parameter. In order to study the evolution of the static sphere radius as the time control parameter, we choose the parameter values $\alpha_1=1,\;\alpha_2/M^2=2.76,\;p/M=0.14,\; q/M=1.02$. For $q\leq q_2$ and $q>q_2$, the metric \eqref{ele} respectively describes a dyonic black hole and naked singularity, with $f(r)$ given in (\ref{fr}). The behavior of the radii of static spheres is shown in Fig. \ref{pprqso}. From the figure, we find that two static spheres emerge only when $q\geq q_1$ for the black hole. The one with small radius has a negative winding number and so is unstable, whereas the one with large radius has a positive winding number and is stable. The total topological number vanishes for $q<q_1$, where static spheres are absent. This property continues for all $q_1<q<q_2$.

At extremality, with $q=q_2$, a new stable static sphere with smaller radius emerges. For $q> q_2$ the solution describes a naked singularity, and there are three static spheres; the topological number is $W$=1-1+1=1, notably different from the situation for a black hole. This situation persists for $q_2<q<q_3$. For $q \to q_{3}$, the two smaller spheres merge, leaving only one static sphere of large radius for $q > q_{3}$. Nevertheless, the topological number $W$=1 remains unchanged. These topological results indicate that if solutions exist, there must be a radially stable static solution, from which a static Dyson sphere can be constructed.

We summarize our results in Table \ref{tab}. The topological number $W$ clearly has distinct values for the black hole and naked singularity, indicating they are in different topological classes.

\begin{table}[]
\setlength{\tabcolsep}{2.5mm}{\begin{tabular}{cccccc}\hline\hline
 & \multicolumn{2}{c}{BH} & &\multicolumn{2}{c}{NS} \\ \cline{2-3}\cline{5-6}
$q$&  $q<q_1$ &  $q_1<q<q_2$ & &    $q_2<q<q_3$  &    $q_3<q$      \\\hline
 $W$ & 0   & 0  & &    1   & 1  \\     \hline\hline
\end{tabular}
\caption{Topological numbers for the black hole and naked singularity for various charges $q$.}\label{tab}}
\end{table}

\section{Summary}
\label{summary}

Quasi-topological electromagnetism provides us with an interesting phenomenological competitor to the standard Maxwell theory, since its basic effects are not manifest in Earth-based laboratory experiments\cite{Weiladf}. Minimally coupling this theory to gravity, we have investigated static spheres, at which a massive particle remains at rest with respect to a static asymptotic static observer in a static, spherically symmetric dyonic black hole and its naked singularity counterpart. This solution respects that that the standard dominant, weak, and null energy conditions are held for positive $\alpha_1$ and $\alpha_2$. We found that the black hole admits a pair of stable and unstable static spheres, and naked singularity admits one more stable one than the unstable ones. They also provide a chance to construct the Dyson-like spheres. What is most intriguing is that, quite unlike standard Einstein-Maxwell theory, no extremality condition is required to achieve static balance.

Making use of a topological argument for static spheres in an asymptotically flat spacetime, we showed the black hole and naked singularity have distinct topological numbers ($W=0$ and 1 respectively) for their static spheres. These results confirm that the stable and unstable static spheres always come in pairs for asymptotically flat black holes. It also indicates that the black hole and naked singularity solutions belong to different topological classes. Moreover, if one the naked singularity solution can be the exterior metric of some stellar structure, there would be a topological phase transition at $q=q_2$ for the static spheres.

Besides realizing the static Dyson-like sphere, the presence of static orbits could result in certain novel gravitational phenomena. For example, there may be one extra static or slow velocity accretion disk, in contrast to those formed at the usual innermost stable circular orbits. Since this orbit, which acts as a light source, is closer to the black hole, it will produce a different shadow pattern. Such double accretion disks will yield interesting observable effects, providing unique tests of general relativity. Although material particles outside may accumulate near the static sphere, their mutual frictional forces and angular momenta will cause these particles to form a metastable accretion disk. An actual disk has a very limited amount of matter, and thus only modifies the local gravitational field, which has very tiny gravitational influence far away from the object inside the sphere. Our main results can be applied to other static, spherically symmetric black holes. As an example, we carried out the calculation for the dyonic black hole with a quasi-topological electromagnetic term, and observed the static orbits. Since quasi-topological electromagnetism is phenomenologically viable, it is conceivable that it replaces Maxwell's theory and that such dyonic black holes might actually exist. Very recently, double-black hole solutions balanced by their scalar hair or cosmic expansion were found exist \cite{Herdeiroh,Dias,Radu}. Our static point orbit naturally provides a mechanism for realizing such double-black hole solutions with the extreme mass ratios. More generally, our study is of particular interest insofar at it provides a deeper understanding of static spheres that might be present in other theories of true observational interest beyond nonlinear electrodynamics.

\section*{Acknowledgements}
This work was supported by the National Natural Science Foundation of China (Grants No. 12075103, No. 12105126, No. 11875151, and No. 12247101), the 111 Project (Grant No. B20063), Lanzhou City's scientific research funding subsidy to Lanzhou University, and the Natural Sciences and Engineering Research Council of Canada.

\end{document}